\documentclass[reprint,superscriptaddress,nofootinbib,twocolumn,amsmath,amssymb,aps,prl,longbibliography]{revtex4-1}

\usepackage{graphicx}
\usepackage{dcolumn}
\usepackage{bm}
\usepackage[colorlinks]{hyperref}

\begin{document}

\preprint{APS/123-QED}

\title{Experimental Implementation of the Fractional Vortex Hilbert's Hotel}

\author{Xi Chen}
\affiliation{Hubei Key Laboratory of Optical Information and Pattern Recognition, Wuhan Institute of Technology, Wuhan 430205, China}

\author{Shun Wang}
\affiliation{Hubei Key Laboratory of Optical Information and Pattern Recognition, Wuhan Institute of Technology, Wuhan 430205, China}

\author{Chenglong You}
\email{cyou2@lsu.edu}
\affiliation{Quantum Photonics Laboratory, Department of Physics \& Astronomy, Louisiana State University, Baton Rouge, LA 70803, USA}

\author{Omar S. Maga\~na-Loaiza}
\email{maganaloaiza@lsu.edu}
\affiliation{Quantum Photonics Laboratory, Department of Physics \& Astronomy, Louisiana State University, Baton Rouge, LA 70803, USA}

\author{Rui-Bo Jin}
\email{jrbqyj@gmail.com}
\affiliation{Hubei Key Laboratory of Optical Information and Pattern Recognition, Wuhan Institute of Technology, Wuhan 430205, China}
\affiliation{Guangdong Provincial Key Laboratory of Quantum Science and Engineering, Southern University of Science and Technology, Shenzhen 518055, China}

\date{\today}

\begin{abstract}
The Hilbert hotel is an old mathematical paradox about sets of infinite numbers. This paradox deals with the accommodation of a new guest in a hotel with an infinite number of occupied rooms. Over the past decade, there have been many attempts to implement these ideas in photonic systems. In addition to the fundamental interest that this paradox has attracted, this research is motivated by the implications that the Hilbert hotel has for quantum communication and sensing. In this work, we experimentally demonstrate the fractional vortex Hilbert's hotel scheme proposed by G. Gbur [Optica 3, 222-225 (2016)]. More specifically, we performed an interference experiment using the fractional orbital angular momentum of light to verify the Hilbert's infinite hotel paradox. In our implementation, the reallocation of a guest in new rooms is mapped to interference fringes that are controlled through the topological charge of an optical beam.
\end{abstract}

\maketitle

\section{Introduction}
The origin of the Hilbert hotel can be traced back to the winter of 1924, when David Hilbert gave a series of lectures on the infinity of mathematics and physics at the University of G\"{o}ttingen \cite{Kragh2014arXiv}.
In these lectures, he used a hotel as an example to illustrate the difference between finite and infinite countable sets. He considered a situation in which a full hotel with a limited number of rooms cannot host new residents. Following this reasoning, he pointed out that a full hotel with an unlimited number of rooms can still host new visitors if every current guest in the hotel is asked to move up one room. This process would lead to a new vacant room, which can be mathematically described with the formula $\infty  + 1 \leftrightarrow  \infty$. As such, this scheme can also be generalized to accommodate any number of guests. Nowadays, this thought experiment is known as the Hilbert's infinite hotel paradox \cite{Kragh2014arXiv}. Remarkably, this paradoxical tale has triggered debate and research in other fields of science including cosmology, philosophy, and theology \cite{Hedrick2014ReligiousStudies,Loke2014ReligiousStudies}.

The Hilbert's hotel paradox has been used to explore the weirdness of infinity, a subject of utmost importance in mathematics. Interestingly, quantum mechanical systems have served as a relevant platform to explore and model the counter-intuitive aspects of this paradox. For example, cavity quantum electrodynamics (QED) platforms have been used to conduct experimental investigations of this paradox \cite{Oi2013PRL}. Specifically, Oi and co-workers experimentally proved that the Hilbert's hotel paradox can be emulated using a cavity QED system with an infinite number of modes. In this case, all the quantum amplitudes were moved up by one level, leaving an unoccupied vacuum state that corresponds to the model of $\infty  + 1 \leftrightarrow \infty$ \cite{Oi2013PRL}. In addition, the orbital angular momentum (OAM) of light has been used to perform experimental implementations of the Hilbert hotel paradox \cite{Potocek2015PRL}. In this regard, Berry predicted the possibility of generating an infinite number of vortex pairs by passing a beam through a half-integer spiral phase plate \cite{Berry2004JOA}.
This prediction was experimentally verified by Leach and colleagues who generated an alternating vortex streamline that confirmed the presence of an infinite number of OAM modes \cite{Leach2004NJP}. However, no transitions in-between the half-integer topological charge was reported. Therefore, one cannot relate this experiment with the verification of the Hilbert's Hotel paradox. Motivated by these ideas, Poto\v{c}ek and colleagues experimentally investigated the Hilbert's hotel paradox by mapping an OAM state to three times the original quantum number \cite{Potocek2015PRL}. More recently, Gbur
identified the possibility of exploiting the propagation of optical beams through fractional vortex plates to perform a direct implementation of the Hilbert's paradox \cite{Gbur2016Optica}. The difference between Poto\v{c}ek's scheme and Gbur's proposal resides on the mechanisms used to accommodate a new guest. It is worth mentioning that the work by Poto\v{c}ek utilizes a quantum state mapping function to represent the room and the guest. In contrast, in Gbur's scheme, the room and the guest are represented by optical vortices with opposite topological numbers. However, Gbur's scheme has not yet been experimentally verified due to the challenges of performing direct measurements of the phase distribution at different spatial locations \cite{OmarPRL2014,MirhosseiniPRL2016, Wang2017OL, Noce2020EPL,OmarReview,Lv2022}.

%
\begin{figure*}[!ht]
\centering
\includegraphics[width=0.85\textwidth]{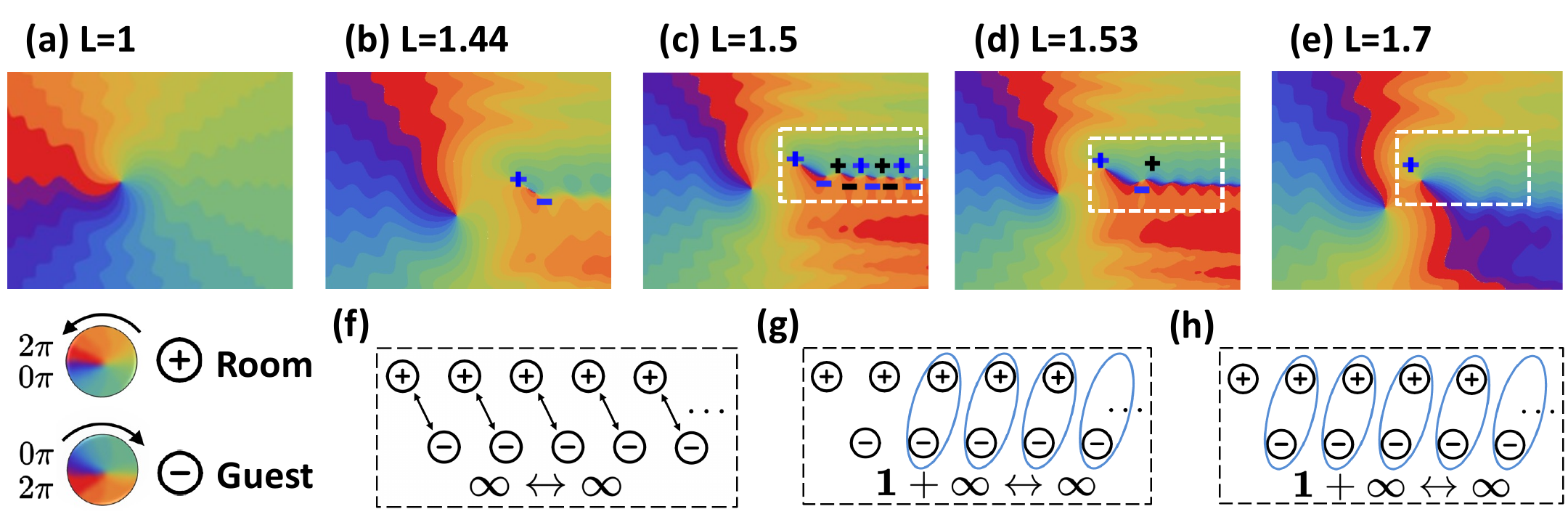}
\caption{The simulated phase distribution of a Gaussian beam after adding a phase. The simulation is performed according to Eq.\,(\ref{eq2}). The corresponding topological charge $L$ in (a-e) are 1, 1.44, 1.5, 1.53, and 1.7, respectively.  The ``+'' sign (blue to red, counterclockwise, positive vortex) indicates a room, whereas the ``-''  sign (blue to red, clockwise, reverse vortex) represents a guest (blue=0, red=2$\pi$). In the panels (f), (g), (h), we illustrate the vortex dynamics simulated in (c), (d), (e).}
\label{fig1}
\end{figure*}
%

Here, we designed an interference experiment using fractional vortex light to verify the Hilbert hotel paradox. Our apparatus enables us to demonstrate the first experimental implementation of the fractional vortex Hilbert's hotel proposed by Gbur \cite{Gbur2016Optica}. By changing the topological charge $L$ from 1.5 to 1.7 in our experiment, we produce different interference structures that enable us to move a guest one room up in the hotel.

\section{Theory}
Following Gbur's proposal \cite{Gbur2016Optica}, we estimate the propagated optical field reflected by a spatial light modulator. For this purpose, we consider a fractional spiral phase $t(\phi )$ encoded on the spatial light modulator (SLM),
\begin{equation}\label{eq1}
t(\phi ) = e^{i\mu \phi},
\end{equation}
here $\mu$ represents the topological charge, which can be any rational number (fractional or integer), and $\phi$ is the azimuth angle. Following the calculation reported in the Supplementary Material \cite{Gbur2017Book, GR07, Zhao2012OE, Li2015, WangLe2019}, it can be shown that the light field at the prorogation distance $z$ and radial coordinate $\rho$ is given by
\begin{equation}\label{eq2}
U_\mu  (\rho ,\phi ,z) = \frac{{\exp \left[ {i\pi \mu } \right]\sin (\pi \mu )}}{\pi }\sum\limits_{n =  - \infty }^\infty  {\frac{{U_n \left( {\rho ,\phi ,z} \right)}}{{(\mu  - n)}}},
\end{equation}
where,
\begin{equation}
    \begin{aligned}\label{eq3}
U_n \left( {\rho ,\phi ,z} \right) = \sqrt {\frac{\pi }{8}}  \exp \left( i k z \right)\exp \left( i n \phi  \right)
\exp \left( {\frac{ik\rho ^2}{4z}} \right)\\
\times \left( { - i} \right)^{\frac{n}{2}} \sqrt {\frac{k\rho ^2 }{z}}
\times \left[ J_{\frac{n-1}{2}} \left( \frac{k\rho ^2 }{4z} \right)  - iJ_{\frac{n+1}{2}} \left( \frac{k\rho ^2 }{4z} \right)  \right].
\end{aligned}
\end{equation}

In this case, $U_n \left( {\rho ,\phi ,z} \right)$ is the propagation field of an integer-order vortex beam, $n$ is an arbitrary integer associated to topological charge, $J_n (x)$ is the Bessel function  of the first kind, $\rho$ is the polar position, and $k$ is the wave number. These parameters enable one to define the  distribution field of $U_\mu  (\rho ,\phi ,z)$. This can be characterized by interfering the field with a plane wave  $E\left( {\rho,\phi ,z} \right)$  \cite{Bazhenov1992, Basistiy1993, Liu2019}. The intensity  distribution associated
with the interference structure can be expressed as
\begin{equation}\label{eq4}
I = \left| {U_\mu  (\rho ,\phi ,z)  + E\left( {\rho,\phi ,z} \right) } \right|^2,
\end{equation}
where
\begin{equation}\label{eq5}
E(\rho ,\phi ,z) = A_0 \exp [ - ik\rho \cos \phi ]\exp [ - \frac{{\rho ^2  - \beta \rho \cos \phi }}{{w^2 (z)}}].
\end{equation}
The parameter $A_0$ describes the amplitude of the plane wave. Furthermore,  $\exp [ - \frac{{\rho ^2  - \beta \rho \cos \phi }}{{w^2 (z)}}]$ defines the range  of the plane wave,  $w (z)$ corresponds to the radius of the beam at position $z$,  and $\beta$ represents a constant related to a horizontal shift.

\section{Simulation}

%
\begin{figure*}[!ht]
\centering
\includegraphics[width=0.85\textwidth]{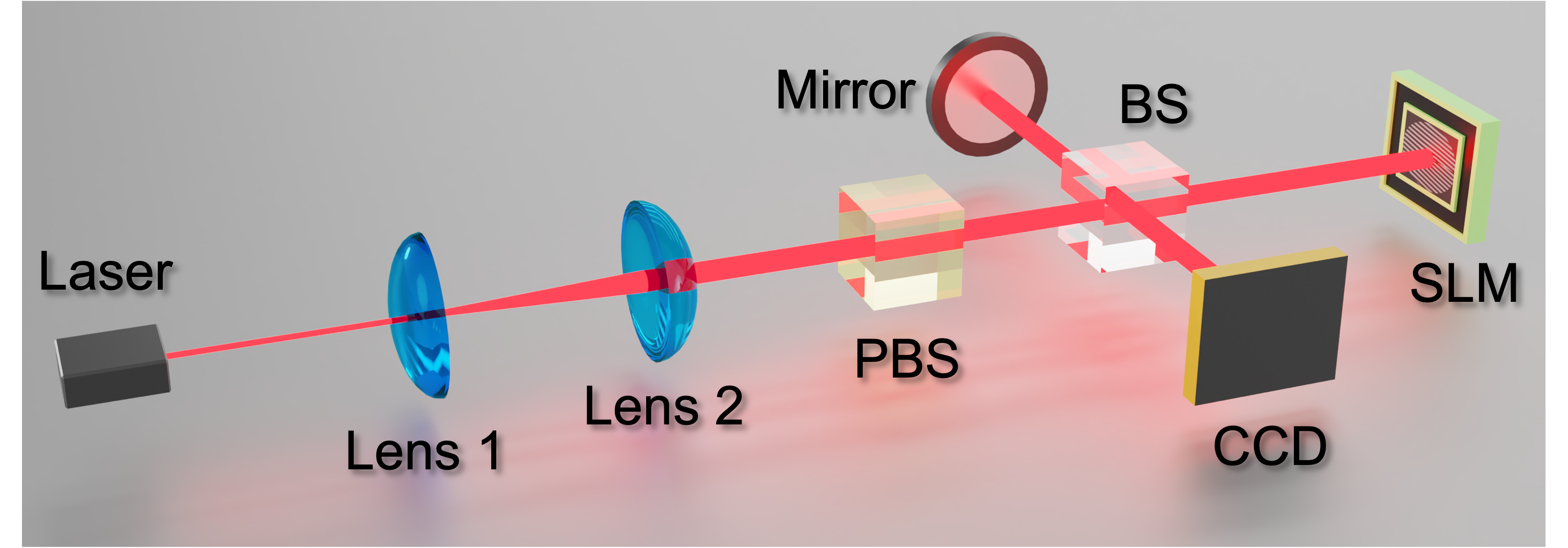}
\caption{The experimental setup. A laser beam at a wavelength of 632.8 nm is expanded by  two lenses (the focal lengths of Lens 1  and Lens 2 are 30 mm and 200 mm, respectively) to  achieve a diameter of 6 mm.  Then, the beam passes through a polarization beam splitter (PBS) to purify the polarization, and injects into a Michelson interferometer, which is composed of a beam splitter (BS), a plane mirror, and a spatial light modulator (SLM). The fork-shaped fringe is loaded onto the SLM to prepare the desired vortex beam. The resulting interference fringes are recorded by a charge-coupled device (CCD) camera.}
\label{fig2}
\end{figure*}
%

Using Eq.\,(\ref{eq2}), we can simulate the propagation of a plane wave reflected by an SLM.
Figure\,\ref{fig1} reports the predictions from our simulation for $\lambda$ = 623.8 nm and $z$ = 0.1 m.
Fig.\,\ref{fig1}(a) corresponds to a situation in which $L = 1$. In this case, we produce a field distribution containing an integer vortex with a phase from 0 to 2$\pi$ rotating clockwise.
Fig.\,\ref{fig1}(b) shows our results for $L=1.44$. This situation produces a fractional vortex field distribution with a gap located to the right. Interestingly,
there is a new pair of vortices formed on the far left of the gap. One rotates clockwise, marked as  ``+'', whereas the other shows a counterclockwise rotation, marked as  ``-''.

 As the value of $L$ increases, the vortex pairs in the opening increase.
When $L=1.5$, as shown in Fig.\,\ref{fig1}(c), the opening  area extends to infinity, and an infinite number of vortex pairs are generated.
This particular situation enables the direct implementation of the Hilbert's infinite hotel model. Namely, a hotel with an unlimited number of rooms, where the rooms are fully occupied by guests.
As illustrated in Fig.\,\ref{fig1}(f), each  ``+'' corresponds to a room, whereas each  ``-'' represents a guest. Thus, room \#1 and guest \#1 are generated together. In general, room \#N and guest \#N are generated together.
Consequently, there are infinite pairs of guests and rooms.
This is equivalent to a one-to-one correspondence between two infinite number sets, namely $\infty   \leftrightarrow  \infty$.

As the value of $L$ continues to increase, for example, from $L = 1.5$ to $1.55$, it is possible to observe that the vortex begins to annihilate at the infinity position.
Specifically, the ``-'' vortex in the previous pair will continuously move to the position of the ``+'' vortex in the back pair.
As a result, the  $N^{th}$ negative vortex and the  $(N+1)^{th}$ positive vortex  are gradually connected,  overlapped, and finally annihilated.
Fig.\,\ref{fig1}(d) shows the situation when $L=1.53$, in this case, the negative vortex is moving close to the positive vortex.
For the  first pair of vortices, there is still some distance between the negative and positive vortices, i.e., there is no annihilation.
However, other infinite vortex pairs have been overlapped and annihilated.
This process corresponds to the model in the Hilbert hotel paradox shown in Fig.\,\ref{fig1}(g). Interestingly, each guest (positive vortex) moves one step to the next room (negative vortex). Here, guest \#1 moves into room \#2, guest \#2  moves into room \#3, and , guest \#N moves into room \#(N+1), while room \#1 on the far left is vacant. In this case, the vortex dynamics represents a one-to-one correspondence between an infinite number set plus one number and an infinite number set, namely $\infty +1  \leftrightarrow  \infty $.

As $L$ increases, the ``-'' vortex in the first pair continues to move to the position of the ``+'' vortex in the second pair. Figure\,\ref{fig1}(e) shows the case in which $L = 1.7$. Under this condition, the positive and negative vortices on the right side have completely overlapped and annihilated. Only an empty room (the positive vortex) remains. The corresponding relationship of  $\infty +1  \leftrightarrow  \infty $ is also  illustrated in Fig.\,\ref{fig1}(h).

\section{Experiment and Results}

%
\begin{figure*}[!ht]
\centering
\includegraphics[width= 0.85\textwidth]{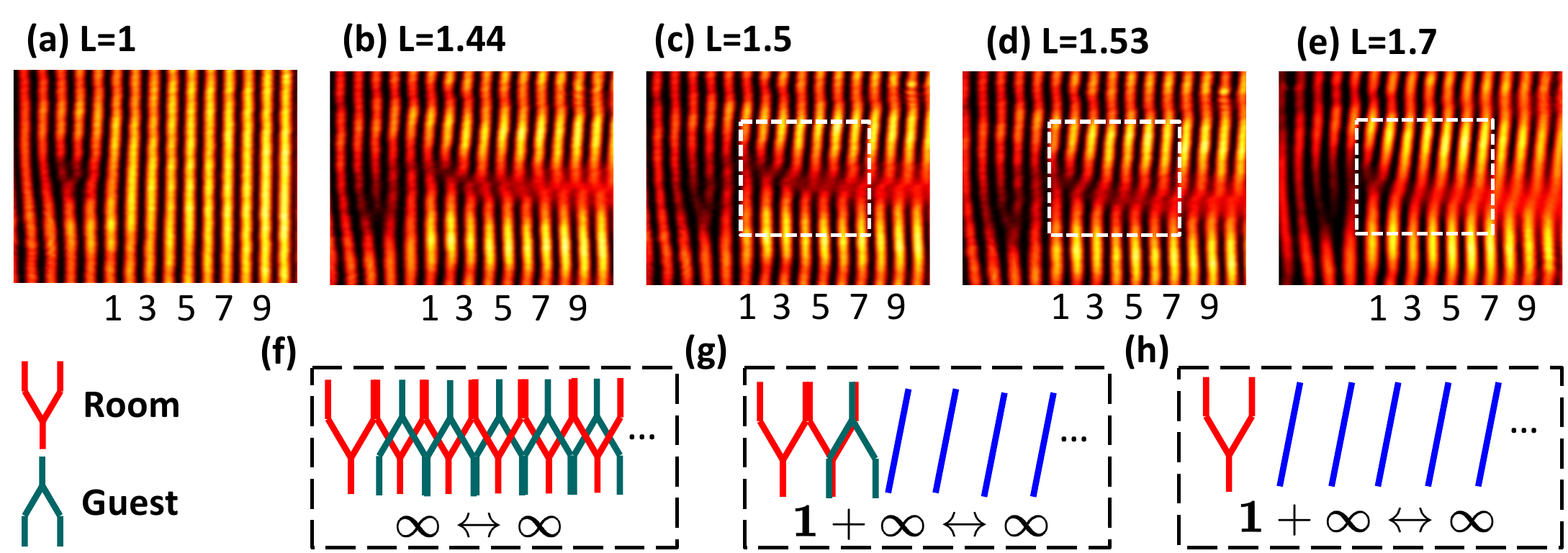}
\caption{Experimental interference fringes obtained by using different values of $L$. The upward-opening-fork shape pattern (red) represents the room, which is obtained by the interference with the positive vortex. The downward-opening-fork shape structure (green) represents the guest, which is obtained by interference with the negative vortex. The tilted blue lines represent tilted interference fringes formed after the overlap and annihilation of positive and negative vortices.}
\label{fig3}
\end{figure*}
%

We have simulated the phase distribution in Fig.\,\ref{fig1} according to Gbur's theory.
However, it is challenging to perform a direct observation of the phase patterns in the experiment \cite{ Zhang:21,PhysRevA.102.023516}. For this reason, we characterize the phase distribution by performing an interference experiment as indicated in Fig.\,\ref{fig2} \cite{Jesus-Silva:12, Wen:19, WEN2020106411, MATTA2020126268, PhysRevA.102.023516}. We expanded
a laser beam at a wavelength of 632.8 nm by using two lenses (focal length L1 = 30 mm; L2 = 200 mm) to  achieve a diameter of 6 mm.
Then, the beam passes through a polarization beam splitter (PBS) that we use to control the initial polarization state. This beam is then injected into a Michelson interferometer, which is composed of a beam splitter (BS), a plane mirror, and a spatial light modulator (SLM, Holoeye, Pluto-2-NIR-011).
The fork-shaped fringe is loaded on the SLM to prepare the desired vortex beam.
Finally, the beam reflected by the SLM interferes with a beam reflected by the plane mirror. In our experiment, we deliberately displaced our interfering beams to observe denser interference patterns. The resulting interference fringes are recorded by a CCD camera (Newport, laser beam profiler, LBP2-HR-VIS2). In Fig.\,\ref{fig3}, we report the experimental interference fringes.

The experimental fork-shaped interference fringes in Fig.\,\ref{fig3} correspond to the simulated phase vortex shown in Fig.\,\ref{fig1}. In particular, Fig.\,\ref{fig3}(a) reports experimental interference fringes measured for $L = 1$.
The original positive vortex in Fig.\,\ref{fig1}(a) changed to an upward-opening-fork shape pattern. Remarkably, more pairs of downward and upward forks are formed as one increases $L$.
Fig.\,\ref{fig3}(b) shows the interference fringes for $L=1.44$. Similarly, the case for $L=1.5$ is shown in Fig.\,\ref{fig3}(c) and illustrated in Fig.\,\ref{fig3}(f). These panels show
countless  upward (red) and downward (green) forked patterns, which are alternately distributed. In this case, we can use the bright upward-opening-fork shape pattern to represent the room, and the bright downward-fork pattern to represent the guest.
This is equivalent to the situation of the Hilbert's infinite hotel model. Indeed, there is a one-to-one correspondence between an infinite number of rooms and an infinite number of guests, namely, $\infty   \leftrightarrow  \infty$.

Figure\,\ref{fig3}(d) and (g) show the case for $L = 1.53$. This condition forces the downward forks (guest) to move close to the upward forks (room).
The downward and the upward-fork patterns slowly overlap to form a long tilted stripe (in blue).
Fig.\,\ref{fig3}(e) and (h) show the case in which $L = 1.7$. Here, guest (downward-fork pattern) \#1   moves into the room  (upward-fork pattern) \#2, guest \#2  moves into the room \#3, and, consequently, guest \#N moves into the room \#(N+1), leaving room \#1 vacant. This relationship can be described as $\infty +1  \leftrightarrow  \infty $.
For sake of completeness, we also performed a theoretical simulation of the interference fringes, see Supplementary Materials S1(c).
It can be seen that the experimental interferogram is in agreement with the simulated interferogram.

\section{Discussions}
Our work reveals that the counter-intuitive concepts associated with infinity can be explored in optical platforms. Furthermore, in the theoretical scheme proposed by Gbur in ref. \cite{Gbur2016Optica}, the Hilbert's infinite hotel paradox was studied using $L = 4.5$. However, our experiment was carried out using $L=1.5$. As such, it should be emphasized that both values enable demonstrating the correspondence $\infty +1  \leftrightarrow  \infty $. The difference resides in the fact that the interference patterns produced for $L=1.5$ are simpler for experimental observations.

In our experiment, we explored the case in which $\infty +1  \leftrightarrow  \infty $. However, this platform can be used to implement the Hilbert's infinite hotel paradox for different correspondences such as $ \infty +N \leftrightarrow  \infty $, or $2 \infty  \leftrightarrow  \infty $. Recently, new theoretical schemes for the Hilbert's hotel paradox were proposed \cite{Wang2017OL, Noce2020EPL}, notably, these schemes may also be experimentally demonstrated using our approach.

\section{Conclusion}
In conclusion, we demonstrated the first experimental implementation of the fractional vortex Hilbert's hotel. In contrast to previous platforms, our scheme enables a direct implementation of the Hilbert's infinite hotel paradox. We believe that our experiment can be used to investigate other counter-intuitive effects associated to infinity.

\section*{Acknowledgments}
We thank Dr. Le Wang (Nanjing University of Posts and Telecommunications) and Dr. Shi-Long Liu (The University of Ottawa) for the helpful discussions. This work was supported by the National Natural Science Foundations of China (Grant Numbers 91836102, 12074299, and 11704290) and by the Guangdong Provincial Key Laboratory (Grant No. GKLQSE202102). C.Y. and O.S.M.L. acknowledge support from Cisco.

\clearpage
\newpage
\onecolumngrid

\renewcommand\thefigure{S\arabic{figure}}
\setcounter{figure}{0}

\setcounter{equation}{0}
\renewcommand\theequation{S\arabic{equation}}
\setcounter{page}{1}

\section{Supplementary Information}

\section*{S1: Calculation of Eq.(\ref{eq3})  }
In this section, we show the derivation of Eq. (\ref{eq3}) in the main text.
\subsection*{S1-1. Calculation of  $A_n (\kappa _x ,\kappa _y )$ at the position of $z=0$ }
In the space domain, at the position of $z=0$, assume there is a light wave with the distribution of $U(x,y)$ on the $xoy$ plane.
\begin{equation}\label{eqS}
U(x,y) = \int_{-\infty }^\infty  \int_{ - \infty }^\infty   A(\kappa _x ,\kappa _y )e^{i(\kappa _x x + \kappa _y y)} d\kappa _x d\kappa _y,
\end{equation}
where $A(\kappa _x ,\kappa _y )$ is spatial frequency distribution  and $\kappa$ is the spatial frequency.
The inverse Fourier transformation of $U(x,y)$ is
\begin{equation}\label{eqS4}
A(\kappa _x ,\kappa _y ) = \frac{1}{{(2\pi )^2 }}\int_{-\infty }^\infty  {\int_{ - \infty }^\infty  {} } U(x,y)e^{ - i(\kappa _x x + \kappa _y y)} dxdy.
\end{equation}
We can change the above formula from a Cartesian coordinate system to a polar coordinate system using
 $x = \rho \cos \varphi$, $y = \rho \sin \varphi$, $\kappa _x  = \kappa \cos \phi _\kappa$ ,  $\kappa _y  = \kappa \sin \phi _\kappa$ and obtain
\begin{equation}\label{eqS5}
A (\kappa ,\phi _\kappa  ) = \frac{1}{{(2\pi )^2 }}\int_0^\infty  {\int_0^{2\pi } \rho U(\rho, \varphi )   } e^{ - i\kappa \rho \cos (\phi  - \phi _\kappa  )}  d\rho d\phi.
\end{equation}
Note in the space domain, $\rho  = \sqrt {x^2  + y^2 }$ and $\phi$ is the polar angle. In the frequency domain, $\kappa  = \sqrt {\kappa _x^2  + \kappa _y^2 } $ and $\phi _\kappa$ is the  polar angle.
For simplicity in the calculation, we assume $U(\rho, \varphi )=1$, i.e., this is a plane wave with each frequency having an equal amplitude of 1.
\begin{equation}\label{eqS7}
A (\kappa ,\phi _\kappa  ) = \frac{1}{{(2\pi )^2 }}\int_0^\infty  {\int_0^{2\pi } \rho  } e^{ - i\kappa \rho \cos (\phi  - \phi _\kappa  )} d\rho d\phi.
\end{equation}

In the space domain, let assume there is a spatial light modulator (SLM) (or other phase component, such as a spiral phase plate) with a transfer function of
\begin{equation}\label{eqS1}
t(\phi ) = e^{i \alpha \phi},
\end{equation}
where $\phi$ is the azimuth angle; $\alpha$  is the phase delay produced by the SLM, and $\mu$  can be an integer or a fraction.
When $n$ is an integer, we get:
\begin{equation}\label{eqS1}
t(\phi ) = e^{i n \phi},
\end{equation}

If the phase $e^{i n \phi}$ is added to the plane wave, we obtain
\begin{equation}\label{eqS8}
A_n (\kappa ,\phi _\kappa  ) = \frac{1}{{(2\pi )^2 }}\int_0^\infty  {\int_0^{2\pi } \rho  } e^{ - i\kappa \rho \cos (\phi  - \phi _\kappa  )} e^{in\phi } d\rho d\phi.
\end{equation}
Let $\phi  - \phi _\kappa  = \phi ^ \prime$, and
\begin{equation}\label{eqS9}
A_n (\kappa ,\phi _\kappa  ) = \frac{1}{{2\pi }}e^{in\phi _\kappa  } \int_0^\infty  {\rho \frac{1}{{2\pi }}\int_0^{2\pi } {e^{ - i\kappa \rho \cos \phi ^\prime } } } e^{in\phi^\prime }  d\rho d\phi ^\prime .
\end{equation}

Considering the standard form of m-order Bessel function:
\begin{equation}\label{eqS10}
J_m \left( x \right) = \frac{1}{{2\pi i^m }}\int_0^{2\pi } {e^{ - im\phi } } e^{ix\cos \left( \phi  \right)} d\phi,
\end{equation}
Eq. (\ref{eqS9}) can be rewritten  as
\begin{equation}\label{eqS11}
A_n (\kappa ,\phi _\kappa  ) = \frac{{i^{\left| n \right|} }}{{2\pi }}e^{in\phi _\kappa  } \int_0^\infty  {J_{\left| n \right|} \left( {\kappa \rho } \right)} \rho d\rho.
\end{equation}
Using the following equation from the standard integral tables (Eq. (14) of Section 6.561 in Ref. \cite{GR07}, also see Ref. \cite{Gbur2017Book}
\begin{equation}\label{eqS12}
\int_0^\infty  {{\rm{x}}^\mu  J_\nu  } \left( {\alpha x} \right)dx = 2^\mu  \alpha ^{ - \mu  - 1} \frac{{\Gamma \left( {\frac{1}{2} + \frac{1}{2}\nu  + \frac{1}{2}\mu } \right)}}{{\Gamma \left( {\frac{1}{2} + \frac{1}{2}\nu  - \frac{1}{2}\mu } \right)}},
\end{equation}
we obtain
\begin{equation}\label{eqS13}
A_{\rm{n}} \left( {\kappa ,\phi _\kappa  } \right) = \frac{{i^{\left| n \right|} }}{{2\pi }}e^{in\phi _\kappa  } 2 \cdot \kappa ^{ - 2} \frac{{\Gamma \left( {\frac{1}{2} + \frac{1}{2}n + \frac{1}{2}} \right)}}{{\Gamma \left( {\frac{1}{2} + \frac{1}{2}n - \frac{1}{2}} \right)}}.
\end{equation}
This equation can be further simplified as
\begin{equation}\label{eqS14}
 A_n \left( {\kappa ,\phi _\kappa  } \right) = \frac{{i^{\left| n \right|} }}{{2\pi }}e^{in\phi _\kappa  } 2 \cdot \kappa ^{ - 2} \frac{{\Gamma \left( {1 + \frac{1}{2}n} \right)}}{{\Gamma \left( {\frac{1}{2}n} \right)}} = \frac{{i^{\left| n \right|} }}{{2\pi }}e^{in\phi _\kappa  } 2 \cdot \kappa ^{ - 2} \frac{{\rm{n}}}{2}.
\end{equation}
The integration result is:
\begin{equation}\label{eqS15}
A_n (\kappa _x ,\kappa _y ) = \frac{{\left| n \right|i^{\left| n \right|} }}{{2\pi \kappa ^2 }}e^{in\phi _\kappa  }.
\end{equation}
Note, in the definition of Gamma function $\Gamma (x)$, $x>0$, therefore, we can change $n$ to be $\left| n \right|$  in the above equation.

\subsection*{S1-2. Calculation of  $U_n \left( {\rho ,\phi ,z} \right)$ at the position of $z$ }
Next,
we consider the distribution of the light field at the position of $z$.
\begin{equation}\label{eqS}
U_{\rm{z}} ({\rm{x}},y) \equiv U({\rm{x}},y,z) = \int_{-\infty}^\infty  {\int_{-\infty}^\infty } A_{\rm{z}} (\kappa _x ,\kappa _y )e^{i(\kappa _x x + \kappa _y y)} d\kappa _x d\kappa _y,
\end{equation}
where $A_{\rm{z}}$ is the distribution of $A$ at the position of $z$.
According to the solution of the Helmholtz equation,
\begin{equation}\label{eqS}
A_{\rm{z}} (\kappa _x ,\kappa _y ){\rm{ = }}A_{\rm{0}} (\kappa _x ,\kappa _y )\exp (i\kappa _z z),
\end{equation}
where $A_{\rm{0}}(\kappa _x ,\kappa _y ) \equiv  A(\kappa _x ,\kappa _y )$ is the distribution of $A$ at the position of $0$.
Therefore, the general distribution of $ U(x,y,z)$ is  (Note, $U(x,y,0)=U(x,y)$)
\begin{equation}\label{eqS2}
U(x,y,z) = \int_{-\infty}^\infty  {\int_{-\infty}^\infty } A(\kappa _x ,\kappa _y )e^{i(\kappa _x x + \kappa _y y + \kappa _z z)} d\kappa _x d\kappa _y,
\end{equation}
The above equation can also be intuitively understood as the diffraction from  U(x,y,0) to U(x,y,z).
In the polar coordinate,  the equation for the case of  $A_n$ can be written as:
\begin{equation}\label{eqS17}
U_n \left( {\rho ,\phi ,z} \right) = \int_0^\infty \int_0^{2\pi}  {A_n } \left( {\kappa _x ,\kappa _y } \right)e^{i\kappa \rho \cos \left( {\phi _\kappa   - \phi } \right)} e^{i\kappa _z z} \kappa d\kappa d\phi _\kappa.
\end{equation}

In order to simplify the calculation, we ignore the evanescent wave and only consider the field propagating near the
z-axis.
 So,
\begin{equation}\label{eqS16}
\begin{array}{l}
 k^2  = \kappa_x ^2  + \kappa_y ^2  + \kappa_z ^2  \\
 \kappa ^2 = \kappa_x ^2  + \kappa_y ^2.  \\
 \end{array}
\end{equation}
Therefore,
\begin{equation}\label{eqS18}
\kappa _z  = \sqrt {{k}^2  - \kappa ^2 }  \approx k - \frac{{\kappa ^2 }}{{2k}}.
\end{equation}

After substituting  Eq.(\ref{eqS18}) into Eq.(\ref{eqS17}), we obtain:
\begin{equation}\label{eqS19}
U_n \left( {\rho ,\phi ,z} \right)  \approx   e^{ikz} \int {A_n } \left( {\kappa _x ,\kappa _y } \right)e^{i\kappa \rho \cos \left( {\phi _\kappa   - \phi } \right)} e^{ - i\kappa ^2 z/2k} \kappa d\kappa d\phi _\kappa.
\end{equation}

After substituting  Eq.(\ref{eqS15}) and Eq.(\ref{eqS10}) into Eq.(\ref{eqS19}), we obtain
\begin{equation}\label{eqS20}
U_n \left( {\rho ,\phi ,z} \right) = e^{ikz} e^{in\phi } \left| n \right|\int_0^\infty  {\frac{{J_{\left| n \right|} \left( {\kappa \rho } \right)}}{\kappa }} e^{ - i\kappa ^2 z/2k} d\kappa.
\end{equation}

In order to simplify the calculation, we set $b=z/2k$ and substitute it into the above formula to get:
\begin{equation}\label{eqS21}
U_n \left( {\rho ,\phi ,z} \right) = \frac{{\rho e^{ikz} e^{in\phi } \left| n \right|}}{2}\int_0^\infty  {\frac{{2J_{\left| n \right|} \left( {\kappa \rho } \right)}}{{\kappa \rho }}} e^{ - ib\kappa ^2 } d\kappa.
\end{equation}

Utilize the standard Bessel equation
\begin{equation}\label{eqS22}
\frac{{2nJ_n \left( x \right)}}{x} = J_{n - 1} \left( x \right) + J_{n + 1} \left( x \right),
\end{equation}
we obtain
\begin{equation}\label{eqS23}
U_n \left( {\rho ,\phi ,z} \right) = \frac{{\rho e^{ikz} e^{in\phi } }}{2}\int_0^\infty  {\left[ {J_{\left| n \right| - 1} \left( {\rho \kappa } \right) + J_{\left| n \right| + 1} \left( {\rho \kappa } \right)} \right]} e^{ - ib\kappa ^2 } d\kappa.
\end{equation}
Using the equation from the standard integral tables
\begin{equation}\label{eqS24}
\int_0^\infty  {{\rm{e}}^{ - \alpha x^2 } J_\nu  } \left( {\beta x} \right)dx = \frac{1}{2}\sqrt {\frac{\pi }{\alpha }} e^{ - \beta ^2 /8\alpha } I_{\nu /2} \left( {\beta ^2 /8\alpha } \right),
\end{equation}
where
\begin{equation}\label{eqS25}
I_m \left( { - ix} \right) = \left( { - i} \right)^m J_m \left( x \right),
\end{equation}
we achieve
\begin{equation}\label{eqS26}
\begin{array}{lll}
 U_n \left( {\rho ,\phi ,z} \right)
&=& \frac{{1 }}{2} \rho e^{ikz} e^{in\phi}   \left\{ {    \int_0^\infty  {\left[ {J_{\left| n \right| - 1} \left( {\rho \kappa } \right)} \right]} e^{ - ib\kappa ^2 } d\kappa {\rm{ + }}\int_0^\infty  {\left[ {J_{\left| n \right| + 1} \left( {\rho \kappa } \right)} \right]} e^{ - ib\kappa ^2 } d\kappa     } \right\} \\
&=&  \frac{{1 }}{2} \rho e^{ikz} e^{in\phi}  \sqrt {\frac{{\pi k}}{{i2z}}  } \times e^{ik\rho ^2 /4z}   \left[ {\left( { - i} \right)^{\frac{{\left| n \right| - 1}}{2}} J_{ \frac{\left| n \right|-1}{2}} \left( { \frac{k\rho ^2 }{4z} } \right) + \left( { - i} \right)^{\frac{{\left| n \right| + 1}}{2}} J_{ \frac{\left| n \right|+1}{2} } \left( {  \frac{k\rho ^2 }{4z} } \right)} \right]. \\
 \end{array}
\end{equation}
After simplification, finally we obtain
\begin{equation}\label{eqS27}
U_n \left( {\rho ,\phi ,z} \right) = \sqrt {\frac{\pi }{8}}  \exp \left( i k z \right)\exp \left( i n \phi  \right)
\exp \left( {\frac{ik\rho ^2}{4z}} \right)
\times \left( { - i} \right)^{\frac{\left| n \right|}{2}} \sqrt {\frac{k\rho ^2 }{z}}
\times \left[ J_{\frac{\left| n \right|-1}{2}} \left( \frac{k\rho ^2 }{4z} \right)  - iJ_{\frac{\left| n \right|+1}{2}} \left( \frac{k\rho ^2 }{4z} \right)  \right].
\end{equation}
If we only consider the case of positive integers, i.e. $\left| n \right|$  can be changed to $n$, then, the above equation is just the Eq. (\ref{eq3}) in the main text.

\section*{S2: Calculation of Eq. (\ref{eq2}) }
In this section, we show how to derive the Eq. (\ref{eq2}) in the main text.
Consider the plane light passing through a fractional-order spiral phase plate.
The transfer function of the fractional-order spiral phase plate can be expanded by an integer order, i.e.,
\begin{equation}\label{eqS}
\exp ( {i\alpha \phi } ) = \frac{{\exp ( {i\pi \alpha } )\sin \left( {\pi \alpha } \right)}}{\pi }\sum\limits_{n =  - \infty }^\infty  {\frac{{\exp ( {in\phi } )}}{{\phi  - n}}}.
\end{equation}
Use Fourier expansion to expand the above formula
\begin{equation}\label{eqS}
\exp ( {i\alpha \phi } )  = \sum\limits_{n =  - \infty }^\infty  {c_k } e^{in\phi },
\end{equation}
where,
\begin{equation}\label{eqS}
\begin{array}{lll}
 c_k  &=& \frac{1}{{2\pi }}\int\limits_0^{2\pi } {e^{i\alpha \phi } e^{ - in\phi } } d\phi  \\
  &=& \frac{1}{{2\pi }}\frac{1}{{i(\alpha  - n)}}e^{i\phi (\alpha  - n)} |_0^{2\pi }  \\
  &=& \frac{1}{{2\pi }}\frac{1}{{i(\alpha  - n)}}[e^{i2\pi (\alpha  - n)}  - 1] \\
  &=& \frac{1}{{2\pi }}\frac{1}{{i(\alpha  - n)}}{\rm{e}}^{i\pi (\alpha  - n)} [e^{i\pi (\alpha  - n)}  - e^{ - i\pi (\alpha  - n)} ] \\
  &=& \frac{1}{{2\pi }}\frac{1}{{i(\alpha  - n)}}e^{i\pi \alpha } e^{ - i\pi n} e^{i\pi n} 2i\sin (\pi \alpha ) \\
  &=& \frac{{\sin (\pi \alpha )}}{\pi }\frac{{e^{i\pi \alpha } }}{{(\alpha  - n)}}. \\
 \end{array}
\end{equation}
Note, in the above calculation,
$ e^{ - i\pi n}  = \cos ( - \pi n) + i\sin ( - \pi n) = \cos (\pi n)$, and
$  e^{i\pi n}  = \cos (\pi n) + i\sin (\pi n) = \cos (\pi n)$, so $e^{ - i\pi n}  = e^{i\pi n} $

Therefore,
\begin{equation}\label{eqS}
e^{i\alpha \phi }  = \frac{{e^{i\pi \alpha } \sin (\pi \alpha )}}{\pi }\sum\limits_{n =  - \infty }^\infty  {\frac{{e^{in\phi } }}{{(\alpha  - n)}}}.
\end{equation}

Following the similar procedure as before, we obtain
\begin{equation}\label{eqS}
\begin{array}{lll}
A_{\alpha} (\kappa ,\phi _\kappa  ) &=& \frac{1}{{(2\pi )^2 }}\int_0^\infty  {\int_0^{2\pi } \rho  } e^{ - i\kappa \rho \cos (\phi  - \phi _\kappa  )} e^{i\alpha \phi } d\rho d\phi \\
&=& \frac{1}{{(2\pi )^2 }}\int_0^\infty  {\int_0^{2\pi } \rho  } e^{ - i\kappa \rho \cos (\phi  - \phi _\kappa  )} \frac{{\exp \left[ {i\pi \alpha } \right]\sin \left( {\pi \alpha } \right)}}{\pi }\sum\limits_{n =  - \infty }^\infty  {\frac{{\exp \left[ {in\phi } \right]}}{{\alpha  - n}}} d\rho d\phi  \\
&=& \frac{{\exp \left[ {i\pi \alpha } \right]\sin \left( {\pi \alpha } \right)}}{\pi }\sum\limits_{n =  - \infty }^\infty  {\frac{1}{{\alpha  - n}}} \frac{1}{{(2\pi )^2 }}\int_0^\infty  {\int_0^{2\pi } \rho  } e^{ - i\kappa \rho \cos (\phi  - \phi _\kappa  )} \exp \left[ {in\phi } \right]d\rho d\phi  \\
&=& \frac{{\exp \left[ {i\pi \alpha } \right]\sin \left( {\pi \alpha } \right)}}{\pi }\sum\limits_{n =  - \infty }^\infty  {\frac{1}{{\alpha  - n}}}  \times A_n (\kappa _x ,\kappa _y ), \\
 \end{array}
\end{equation}
where,
\begin{equation}\label{eqS}
A_n (\kappa _x ,\kappa _y ) = \frac{{\left| n \right|i^{\left| n \right|} }}{{2\pi \kappa ^2 }}e^{in\phi _\kappa  }.
\end{equation}
Change from frequency domain to space domain,
\begin{equation}\label{eqS}
\begin{array}{lll}
U_\alpha  \left( {\rho ,\phi ,z} \right) &=& \frac{{\exp \left[ {i\pi \alpha } \right]\sin \left( {\pi \alpha } \right)}}{\pi }\sum\limits_{n =  - \infty }^\infty  {\frac{1}{{\alpha  - n}}}  \times e^{ikz} \int {A_n } \left( {\kappa _x ,\kappa _y } \right)e^{i\kappa \rho \cos \left( {\phi _\kappa   - \phi } \right)} e^{ - i\kappa ^2 z/2k} \kappa d\kappa d\phi _\kappa\\
&=& \frac{{\exp \left[ {i\pi \alpha } \right]\sin \left( {\pi \alpha } \right)}}{\pi }\sum\limits_{n =  - \infty }^\infty  {\frac{1}{{\alpha  - n}}}  \times U_n \left( {\rho ,\phi ,z} \right),\\
 \end{array}
\end{equation}
Finally, we obtain
\begin{equation}\label{eqS}
U_\alpha  (\rho ,\phi ,z) = \frac{{\exp \left[ {i\pi \alpha } \right]\sin (\pi \alpha )}}{\pi }\sum\limits_{n =  - \infty }^\infty  {\frac{{U_n \left( {\rho ,\phi ,z} \right)}}{{(\alpha  - n)}}}.
\end{equation}
This is just the Eq. (\ref{eq2}) in the main text.

\section*{S3: Calculation of Eq. (\ref{eqS18}) }
Here we derive Eq. (\ref{eqS18}):
\begin{equation}\label{eqS}
\sqrt {k^2  - \kappa ^2 }  \approx k - \frac{{\kappa ^2 }}{{2k}}.
\end{equation}
According to the Newton's generalized binomial theorem:
\begin{equation}\label{eqS}
\begin{array}{lll}
\left( {x + y} \right)^r  &=& \sum\limits_{k = 0}^\infty  {\left( {\begin{array}{*{20}c}
   r  \\
   k  \\
\end{array}} \right)} x^{r - k} y^k  \\
&=& x^r  + rx^{r - 1} y + \frac{{r(r - 1)}}{{2!}}x^{r - 2} y^2  + \frac{{r(r - 1)(r - 2)}}{{3!}}x^{r - 3} y^3  + ...,\\
 \end{array}
\end{equation}
we obtain,
\begin{equation}\label{eqS}
\sqrt {k^2  - \kappa ^2 }  = (k^2  - \kappa ^2 )^{1/2}  = \left( {k^2 } \right)^{1/2}  + \frac{1}{2}\left( { - k^2 } \right)^{ - 1/2}  \cdot \kappa ^2 + O \approx k - \frac{{\kappa ^2 }}{{2k}}.
\end{equation}

\section*{S4: Comparison of the simulation and experimental results}
\begin{figure}[!b]
\centering
\includegraphics[width= 0.95\textwidth]{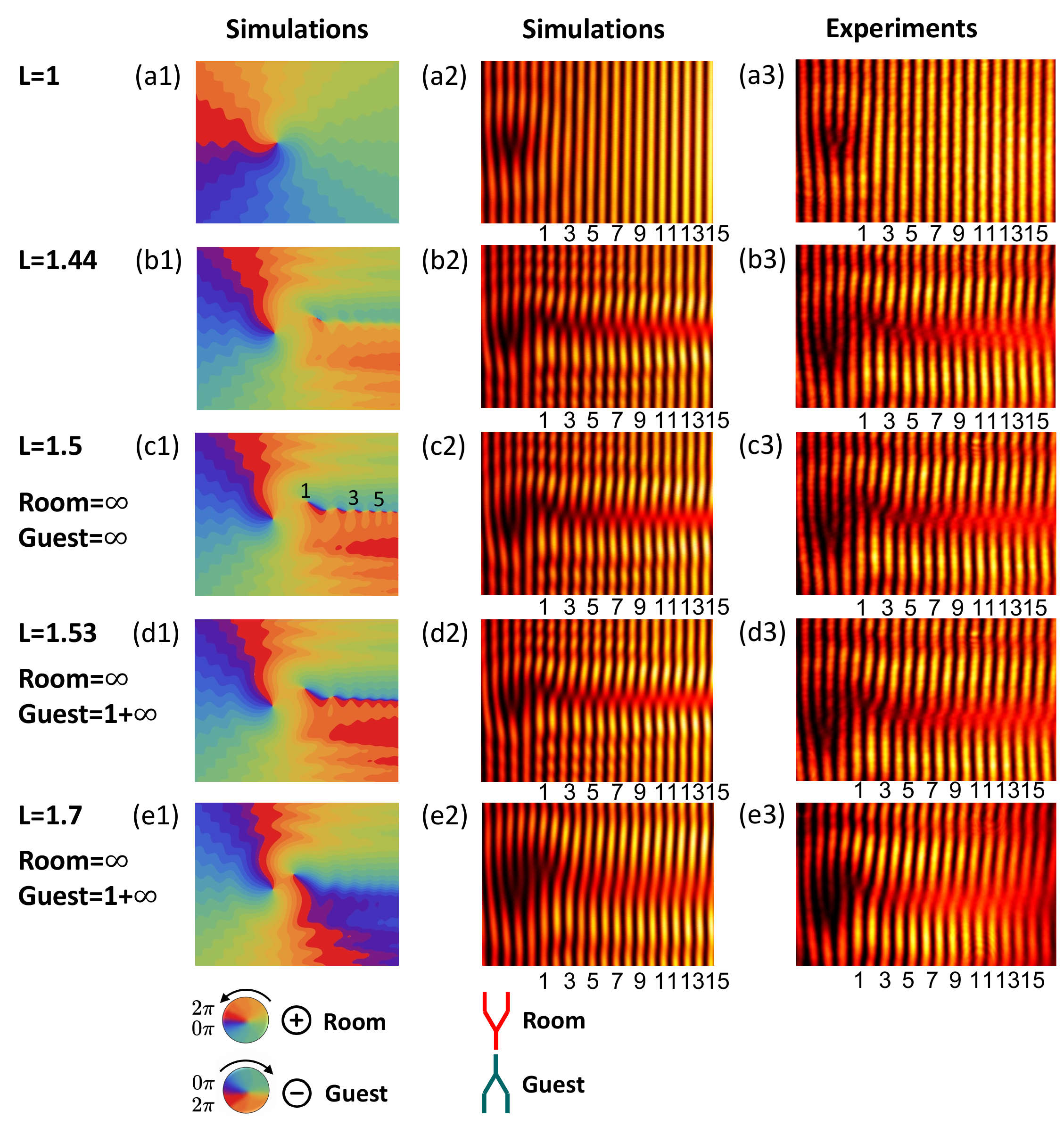}
\caption{ Comparison of the numerical simulations and experimental results at different L values of 1, 1.44, 1.5, 1.53, and 1.7: (a1-e1) The simulations of phase distribution by using Eq.\,(\ref{eq2}); (a2-e2) The simulations of interference patterns by using Eq.\,(\ref{eq4}); (a3-e3) The experimental interference patterns.}
\label{fig-S1}
\end{figure}

In Fig.\,\ref{fig-S1}, we compare the phase simulation graphs by using Eq.\,(\ref{eq2}),  the interference graph simulated by using Eq.\,(\ref{eq4}), and  the interference experiment results.  It can be seen that the theoretical simulation of interference in Fig.\,\ref{fig-S1}(b) corresponds well to the experimental interference patterns in Fig.\,\ref{fig-S1}(b) .
%
%

\end{document}